\documentclass{aa}
\usepackage{graphicx}
\usepackage{txfonts}
\usepackage{amssymb}

\begin{document}
   
   \title{On the correlation between the X-ray and gamma-ray emission in
          TeV blazars}
      
   \author{Katarzy\'nski K. \inst{1} \& 
           Walczewska K.    \inst{1}
	  }

   \offprints{Krzysztof Katarzy\'nski \\kat@astro.uni.torun.pl}

   \institute{
              $^1$Toru\'n Centre for Astronomy, Nicolaus Copernicus University, 
              ul. Gagarina 11, PL-87100 Toru\'n, Poland
             }

   \date{Received 18 August 2009 / Accepted 17 November 2009}

   \abstract{}
            {The observations of TeV blazars published recently show an unexpected 
             quadratic or even cubic correlation between the X-ray and gamma-ray 
             emission. A standard model of the synchrotron self-Compton emission of 
             a compact source inside a jet is not able to explain such a correlation. 
             Therefore, we propose an alternative scenario where the emission of 
             at least two independent compact components is observed at the 
             same time.
            }
            {We compare two different models. 
             The first model assumes the injection of relativistic particles into a downstream 
             region of a shock wave inside a jet that creates the emitting source. The model
             precisely describes the evolution of the particle energy spectrum inside the
             source and takes into account a light-crossing time effect for the produced
             radiation. The second model assumes an intrinsically constant emission of 
             a homogeneous source that travels inside the jet along a curved trajectory,
             where the activity is produced simply by different values of the
             source's Doppler factor. To verify the two models we use recentlu published
             observations of Mrk 421.
            }
            {Our simulations show that simultaneous radiation of at least two 
             independent sources, where the first source dominates the emission in 
             the X-ray range and the second source radiates strongly in the 
             gamma-ray range, can explain the observed correlations. 
             However, the injection model provides inadequate results because 
             it gives different values for the correlation of the rise and 
             decay of a flare. This problem is negligible in the scenario that
             uses the Doppler boosting effect. Therefore, this approach yields
             much better results.
            }
            {}

\titlerunning{On the correlation...}
\authorrunning{Katarzy\'nski et al.}   
\maketitle

\section{Introduction}

Activity of TeV blazars is usually observed simultaneously in the
X-ray and the gamma-ray range (e.g. Catanese et al. \cite{Catanese97},
Pian et al. \cite{Pian98}, Sambruna et al. \cite{Sambruna00},
Takahashi et al. \cite{Takahashi00}, Krawczynski et al. 
\cite{Krawczynski02}, Donnarumma et al. \cite{Donnarumma09}). 
Therefore, the activity can be analysed by a simple comparison 
of the two different light curves. Such a comparison has shown in 
a few cases a surprisingly precise correlation between the evolution
of the X-ray and the gamma-ray emission. This provides an excellent 
opportunity to test theoretical models for very high energy 
emission of blazars.

To describe the correlation is convenient to assume that the
evolution of the X-ray or gamma-ray flux during an outburst can 
be approximated by a power-law function ($F_{\rm X} \propto 
t^s$ and $F_{\rm TeV} \propto t^c$) with different values of $s, c$
for the rise and decay phase of the activity. Rewriting the first 
relationship as $t \propto F_{\rm X}^{1/s}$ and substituting 
this to the second proportionality we can define the 
correlation
\begin{equation}
F_{\rm TeV} \propto F_{X}^{c/s},
\end{equation}
which is also a power-law function with the index $x=c/s$. Such a simple 
definition can describe precisely the correlations observed in TeV 
blazars.

The analysis of the light curves produced by different blazars shows
that there is no unique value of the correlation slope. 
The activity of Mrk 501 observed in April 1997 (Catanese et al. 
\cite{Catanese97}) gives  $x = 1.71 \pm 0.50$ for the comparison 
between the 50-150 keV light curve obtained by the OSSE experiment 
and the gamma-ray observations obtained above 350 GeV by the Whipple 
telescope. On the other hand the comparison between the 2-10 keV 
observations made by the RXTE-ASM experiment and the gamma rays 
detected by the Whipple for the same activity of Mrk 501 gives 
$x=2.69 \pm 0.56$ (Katarzynski et. al. \cite{Katarzynski05}).
Moreover, observations of Mrk 501 conducted by the RXTE-PCU 
experiment (2-20 keV) and the gamma-ray telescopes
(HEGRA and Whipple) show a linear correlation ($x=0.99 \pm 0.01$) 
for the data obtained in May 1997 and an almost quadratic relation 
($x=2.07 \pm 0.27$) for the observations made in June 1998 
(Gliozzi et al. \cite{Gliozzi06}).

The observations of Mrk 421 made in March 2001 (Fossati et al. 
\cite{Fossati08}) show several outbursts observed simultaneously 
in the X-rays (RXTE) and the gamma rays (HEGRA, Whipple). The detailed 
data analysis performed by the authors of the observations shows at 
least in one case more than a quadratic correlation ($x > 2$) and 
in two cases significantly more than linear relation ($x > 1$).

Recent observations of PKS 2155-304 made by $Chandra$ and the H.E.S.S. 
experiments show a much more steeper correlation with the index $x \simeq 3$
(Acharonian et al. \cite{Aharonian09}). However, in this particular case
the correlation was observed mostly in a decay phase of the flare because
of a delay of the {\it Chandra} observations. Steep correlation means a
relatively small change of the X-ray flux in comparison with the variation
of the gamma-ray emission. This means that in an extreme case, where $x \to 
\infty$ only gamma-ray activity can be observed. This was reported 
at least two times (e.g. Krawczynski et al. \cite{Krawczynski04}, Blazejowski
et al. \cite{Blazejowski05}) and is know in the literature as an {\it orphan}
flare phenomenon.

The correlation can be well determined only when an activity event 
is well observed by two different instruments. This requires very good
sampling of the recorded light curves (at least several observations 
during a flare) and small error values in comparison with the amplitude
of the variations. These conditions made it possible for only a few
cases to precisely determine the index of the correlation so far.
So was the correlation sufficiently determined only for three out of
nine cases analysed by Fossati et al. (\cite{Fossati08}). Also more recent 
observations of TeV blazars do not give a definitive answer about the 
correlation slope (e.g. Albert et al. \cite{Albert07a}, Horan et al. 
\cite{Horan09}, Bonnoli et al. \cite{Bonnoli09}). 

The observations show that the correlation slope is changing form
linear to cubic. However, the correlations obtained for a relatively long 
period of the observations (weeks or moths) are usually 
linear or slightly more than linear (e.g. Gliozzi et al. \cite{Gliozzi06},
Albert et al. \cite{Albert07b}), whereas the observations of short
flaring events (a few hours) give quadratic or even cubic relations
(e.g. Fossati et al. \cite{Fossati08}, Aharonian et al. \cite{Aharonian09}).
Moreover, a scatter of the correlated data points seems to be much
higher for the long period correlations in comparison with the
results obtained for the short events. The long time correlation contains 
observations of many different flares produced by different components 
of a jet, and this is probably the reason of the scatter. Therefore, the
short time correlations should provide much better constraints for the
emission models. Especially interesting are cases where the short time
correlation is quadratic or more than quadratic. Standard one-zone 
models frequently used to explain the high energy emission of blazars
are not able to explain such a slope of the correlation 
(Katarzynski et al. \cite{Katarzynski05}). In the present 
work we propose a more complex approach, where the emission of at least 
two independent sources is observed at the same time.

\section{The problem of the quadratic correlation}

The most simple model that is able to explain the high energy emission of TeV
blazars assumes a compact source located inside the jet at a distance of
less that 1 pc from the centre. The source is filed uniformly by 
relativistic electrons and a tangled magnetic field. The particles spinning
around the magnetic field lines are producing a synchrotron emission which
is usually observed in the X-ray range. Some fraction of this emission
is up-scattered to higher energies by the same population of the
electrons. This is the well-known synchrotron self-Compton (hereafter SSC)
radiation that appears in the gamma-ray range. This simple scenario
was used many times to explain the high energy spectra of TeV blazars
(e.g.  Bloom \& Marscher \cite{Bloom96}, Ghisellini et al. \cite{Ghisellini96}, 
Inoue \& Takahara \cite{Inoue96}, Mastichiadis \& Kirk \cite{Mastichiadis97},
Krawczynski et al. \cite{Krawczynski00}, Katarzynski et al. \cite{Katarzynski01}).

The intensity of the synchrotron emission is proportional to 
the particle density, whereas the intensity of the SSC radiation 
is proportional to the square of the particle density. This 
well-known relationship could explain in principle the observed 
quadratic correlations if only the change of the particle density 
were responsible for the observed activity. The question is
how realistic such a scenario is.

An injection of the relativistic particles into the source, which
increases the density, could in principle explain the quadratic 
correlation during the rising phase of a flare. But the particles 
should be injected simultaneously into the entire volume of the source. 
Moreover, the source volume and the magnetic field inside the source
should remain constant during the injection phase. The radiative 
cooling of the particles should also be negligible during the injection.
All these requirements render this scenario unrealistic. We will
demonstrate that a more realistic scenario, which assumes a local 
injection into an expanding source where radiative cooling is important, 
leads to linear correlation during the rising phase.

By analogy systematic energy-independent escape 
of the particles which decrease the density could in principle explain the
quadratic correlation during the decay phase of a flare. This process
requires a significantly weaker magnetic field outside the source
in order to reduce the efficiency of the synchrotron emission. On the other 
hand the radiation field energy density outside a spherical ($R$ - radius)
homogeneous source at a distance of $1/2 R$ above the source surface is
only half as weak as on the surface (Gould \cite{Gould79}).
This means that particles outside the source can still efficiency 
produce gamma rays through the inverse-Compton scattering. 
In other words, the gamma-ray emission will not decay fast enough 
to produce the quadratic correlation during the decay phase.

The detailed analysis of the correlation for many different
scenarios of a single source evolution was performed by Katarzynski 
et al. \cite{Katarzynski05}, where simple analytic formulae were 
derived to describe basic cases, and more complex scenarios were 
analysed through numerical simulations. This analysis shows that
in all realistic cases we should expect rather linear than 
quadratic correlation. 

Finally, in the case of PK2155-304 the correlation with the index
$x \simeq 3$ was observed  (Aharonian et al. \cite{Aharonian09})
and this certainly cannot be explained by the changes of the 
particle density alone.

\section{Single source vs two sources}

A steep slope of the correlation ($x \gtrsim 2$) can be easily obtained if 
we consider simultaneous emission of at least two sources. But let us first
assume the emission of a single source, where the X-ray 
flux is increasing during a rising phase of a flare as a 
power-law function, where
\begin{equation}
F_{\rm X, min} = a~t_{\rm min}^s, ~~~~~F_{\rm X, max} = a~t_{\rm max}^s
\end{equation}
is the flux at the beginning ($t_{\rm min})$ and the maximum ($t_{\rm max}$) 
of the activity respectively. Comparing the two fluxes we can define 
the ratio of the X-ray fluxes in the rising phase
\begin{equation}
F_{\rm X,r} = \frac{F_{\rm X, max}}{F_{\rm X, min}} = \frac {t_{\rm max}^s}{t_{\rm min}^s},
\end{equation}
and derive the power-law index
\begin{equation}
s = \frac{\ln(F_{\rm X,r})}{\ln(t_{\rm max})-\ln(t_{\rm min})}.
\end{equation}
We can repeat this assumption and calculation for the
gamma-ray emission, which gives the index for the flux evolution in
a similar form
\begin{equation}
c = \frac{\ln(F_{\rm TeV,r})}{\ln(t_{\rm max})-\ln(t_{\rm min})}.
\end{equation}
Therefore, the slope of the correlation can be described by a
simple formula
\begin{equation}
x = \frac{c}{s} = \frac{\ln(F_{\rm TeV, r})}{\ln(F_{\rm X, r})}
\label{equ_corr_slope}
\end{equation}
that is valid for a single-source emission. 

The formula derived above shows that the correlation slope can
be increased by an increase of the gamma-ray flux ratio 
($F_{\rm TeV, r}$) or by a decrease of the X-ray flux radio 
($F_{\rm X, r}$). The first option is limited by the nature 
of the SSC emission which gives at maximum a quadratic relation 
in the specific conditions that are rather not realistic, 
as we discussed already. We decided accordingly to test
the second possibility and to decrease the X-ray flux ratio.
This ratio will decrease significantly when we assume an
additional source of the X-ray emission. In the simplest
case we can assume a constant level of this additional 
emission in a time $F_{\rm X,~const}$. This gives
\begin{equation}
F'_{\rm X, r} = \frac{F_{\rm X, max} + F_{\rm X,~const}}{F_{\rm X, min} + F_{\rm X,~const}} \leqslant 
               \frac{F_{\rm X, max}}{F_{\rm X, min}} = F_{\rm X,r}
\end{equation}
for $F_{\rm X,~const} \geqslant 0$ and shows that $F_{\rm Xq,~const}$ 
must be comparable or greater than $F_{\rm X, min}$ to significantly
modify the ratio. On the other hand, possible gamma-ray emission of 
this additional source should be negligibly small in comparison with
the first source emission to avoid simultaneous modification of 
$F_{\rm TeV, r}$. Moreover, with additional X-ray emission the
correlation is no longer a simple power-law function and Eq.
\ref{equ_corr_slope} can be used only for the estimation of 
the correlation slope. But the derived formulae explain
why the correlation slope can be steeper also in more complex
situations which are difficult to describe using simple 
analytic relations.
%}

\section{The observations}

In this work we focus on the activity of Mrk 421 observed during the 
campaign of observations conducted in March 2001 by RXTE, HEGRA and
the Whipple experiments (Fossati et al. \cite{Fossati08}). The seven days 
long light curves obtained during this campaign were divided into nine 
periods, and the activity events were analysed separately. The best 
results were obtained for the first, fourth and fifth night of the 
observations (March 18/19, 21/22, 22/23). 

More than quadratic correlation was observed only during the first night. 
The comparison of the X-ray flux (0.2-10 keV) and all the available 
TeV observations ($E>0.4$ TeV) give in this case $x=2.26 \pm 0.25$. 
But the comparison between the X-ray data and the TeV observations 
made only by Whipple during this night gives the almost cubic value of 
the correlation $x=2.84 \pm 0.41$. What is important, only Whipple 
was observing a strong flare during this night, the light curve 
obtained by HEGRA before the flare shows no activity at all.

The correlation obtained for the fourth and fifth night was
less than quadratic but significantly more than linear with 
$x=1.56 \pm 0.25$ and $x=1.67 \pm 0.16$ respectively. Note that 
the relative amplitudes of the flux changes are significantly 
lower in these observations.

The flare observed during the first night seems to be one of the 
best activity events ever observed in TeV blazars. The evolution 
of the X-ray and gamma-ray flux was well detected during the 
rise and the decay phase of the flare. This gives very important 
information about the correlation which seems to be very similar
for the rising and decay phase. Moreover, the excellent observations
give also information about the spectral evolution of the 
emission. Finally, this activity seems to be quite simple, created 
by one or two sources, whereas the light curves obtained during 
the other nights looks like a superposition of many events in time. 
We consequently selected this particular flare to test our models.
If a model is able to explain the correlation obtained for this
extreme flare, it should be able to explain any other correlation
as well.

\section{Injection scenario}

To explain the correlation it is necessary to apply time-dependent 
modeling. Many different models have been proposed to explain the
evolution of the high energy activity in TeV blazars so far (e.g. 
Dermer \cite{Dermer97}, Kirk et al. \cite{Kirk98}, Coppi, \& Aharonian 
\cite{Coppi99}, Kataoka \cite{Kataoka00}). Most of them assume
changes in the particle energy spectrum inside a source as the
main reason of the activity. Some more complex models additionally 
are taking into account the light-crossing time effect 
(e.g. Chiaberge \& Ghisellini \cite{Chiaberge99}, Sokolov et al.
\cite{Sokolov04}, Graff et al. \cite{Graff08}). 

In the first part of this work we use the relatively simple model 
proposed by Chiaberge \& Ghisellini (\cite{Chiaberge99}) and 
improved by Katarzynski et al. (\cite{Katarzynski08}). The model
assumes particle acceleration by a shock wave inside a jet. The
shock is created by two or more colliding components of the jet,
where some fraction of the kinetic energy of the components is
converted into random energy of the particles. This is the 
well-known internal shock scenario proposed for the first time by
Rees (\cite{Rees78}). However, there is no precise description 
of the acceleration process in the model we use. The particle 
energy increases relatively fast at the shock region and 
after that most of the particles escape into the downstream 
region of the shock. Therefore, the acceleration can be 
approximated as an injection of the particles into 
some volume. In this particular case we assume a power-law
energy distribution of the injected particles 
$Q(\gamma)=Q_{\rm inj} \gamma^{-n}$, where the particle energy
is given by $E=\gamma m_e c^2$. The geometry of the shock front
is square with the length $R$ and the thickness $\ll R$. The duration 
of the injection process is $t=R/c$. The injection 
forms a cube like source. This particular geometry was chosen to 
describe the light-crossing time effect. The source volume was
divided into $10 \times 10 \times 10$ cells. Summing the emissions
of the cells in the proper way (detailed description in  
Chiaberge \& Ghisellini \cite{Chiaberge99}) makes it possible to
simulate this effect.

As we already discussed, a single zone model is not able to explain a
quadratic or a more than quadratic correlation. Therefore we propose 
more complex scenario, where the X-ray emission is produced mostly
by a relatively large source ($R \sim 10^{16}$ cm). This source has a
relatively small particle density and is therefore not able to produce 
efficient gamma-ray emission. To explain TeV gamma rays we use another 
source that is compact ($R \sim 10^{15}$ cm) and dense, and 
is able to dominate the emission in this energy range. 
The two sources are completely independent and located at different
positions inside the jet. But by chance the emission of both
sources is observed at the same time. For simplicity we use the same
injection scenario, described above, to calculate the evolution of both 
sources. 

Note that the scenario which assumes the simultaneous emission of two
or more sources was already successfully used to explain the rapid
TeV variability observed in PKS 2155-304 (Aharonian et al. 
\cite{Aharonian07}). Compact sources (of a size of about $10^{14}$ cm) 
which produce TeV emission several times stronger than X-ray radiation 
can explain activity, where the variability time scale is of about 
a few minutes. The main problem of compact and dense sources --
absorption of TeV emission due to electron-positron par production, 
is negligible in such an approach (Katarzynski et al. 
\cite{Katarzynski08}).

\begin{figure}[!t]
\resizebox{\hsize}{!}{\includegraphics{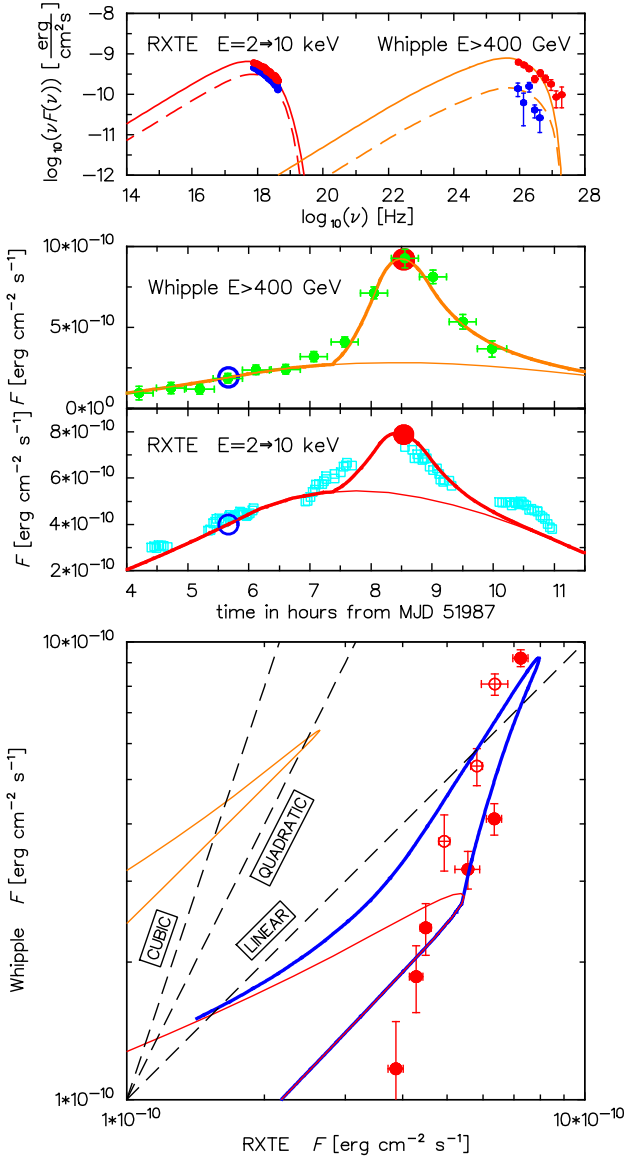}}
\caption{The activity of Mrk 421 observed by Whipple and RXTE experiments on
         19 of March 2001 (Fossati et al.\cite{Fossati08}) and the results of 
         our modeling which assumes the particle injection. The upper panel shows 
         spectra obtained for two different time moments: before the main 
         flare and at the top of the flare. The middle panels show the light
         curves where the total emission is indicated by the thick lines and,
         the thin lines show only the extended source emission. The dots in 
         the middle panels show the time moments, where the spectra were obtained.
         Lower panel show the observed correlation and the results of the 
         modeling, where the open symbols indicate the decay phase and the 
         thin lines show the correlations calculated separately for each source.
        }
\label{fig_inj}        
\end{figure}

The results of our modelling are presented in Fig. \ref{fig_inj}, where
we show spectra selected for two different time moments, the light
curves and the obtained correlation. The source size determines directly 
the variability time scale. Accordingly, a more extended source produces
relatively slow variability, whereas the compact source explains
the flare. The correlation obtained from the modelling does not
perfectly reproduce the observed relation. The correlation index is
different for the rising and decay phase of the activity. This is
directly related to the fact that the increment of the emission
is produced by the injection, whereas the decay is caused by the
radiative cooling. The two completely different physical processes
control the evolution during the rise and decay of the activity and
the correlation has also different slopes during these
phases. In Fig. \ref{fig_inj} we show the total correlation obtained
for both sources simultaneously and the correlations calculated 
separately for each source. The correlation calculated for the
single source has a linear slope ($x_{\uparrow} \simeq 1$) during 
the rising phase and and an almost square root slope ($x_{\downarrow} 
\simeq 0.5$) in the decay phase. The total correlation shows that
the slopes are changing significantly when the emission of two sources
is observed simultaneously ($x_{\uparrow} \simeq 1 \to 3$ and 
$x_{\downarrow} \simeq 0.5 \to 2 $). This helps to explain the
observed correlation, but the difference between $x_{\uparrow}$
and $x_{\downarrow}$ produced within the single source appears
also in the total correlation. 

The values of the physical parameters used in the modelling which are
identical for both sources are: Doppler factor $\delta=20$,
$\gamma_{\rm min} = 1$, $\gamma_{\rm max} = 10^6$, $n = 2$. The difference
appears in the source size $R= 1.1 \times 10^{16}$ and $1.77 \times 10^{15}$ cm, 
the magnetic field strength $B=0.05$ and $0.08$ G and the density of the
injected particles $t_{\rm inj} Q_{\rm inj} = 1.7 \times 10^4$ and $ 
7 \times 10^5$ cm$^{-3}$ for the extended and the compact source 
respectively. The number of free parameters shows that the model
we use is quite simple, and this was the main reason to chose this
particular scenario. A detailed description of the model is given
in Chiaberge \& Ghisellini (\cite{Chiaberge99}) and Katarzynski 
et al. (\cite{Katarzynski08}). 

Note that we have tested also a scenario which precisely describes the 
acceleration process (Katarzynski et al. \cite{Katarzynski06}). This 
particular model gives a different value of the correlation for 
a single source emission. However, in this case $x_{\uparrow}$ 
is also significantly different from $x_{\downarrow}$ because two different
processes (acceleration and cooling) control the rise 
and decay of a flare. This appears to be the main problem for the 
models which are trying to explain the activity by the evolution of 
the particle energy spectrum. It seems that within all scenarios 
which assume the particle energy evolution, only adiabatic 
compression and then adiabatic expansion of the source could 
give $x_{\uparrow} = x_{\downarrow}$. But, this approach
requires a negligible radiative cooling which makes this scenario 
not realistic. 

\section{Doppler boosting effect}

Simultaneous emission of two or more sources can explain any slope of the
correlation where the slope depends on the relative value of the X-ray
and the gamma-ray emission of the sources. However, the observed 
correlation seems to be very similar for the rising and decaying phase
of a flare. This appears to be problematic for the models that
assume different physical processes to explain the two phases of
the activity. We decided to test a very simple scenario
that assumes changes of the Doppler factor as a main reason of the 
flux variations. This idea was already proposed several times -- for
example to explain the activity of Mrk 501 (Villata \& Raiteri \cite{Villata99}).

\begin{figure}[!t]
\resizebox{\hsize}{!}{\includegraphics{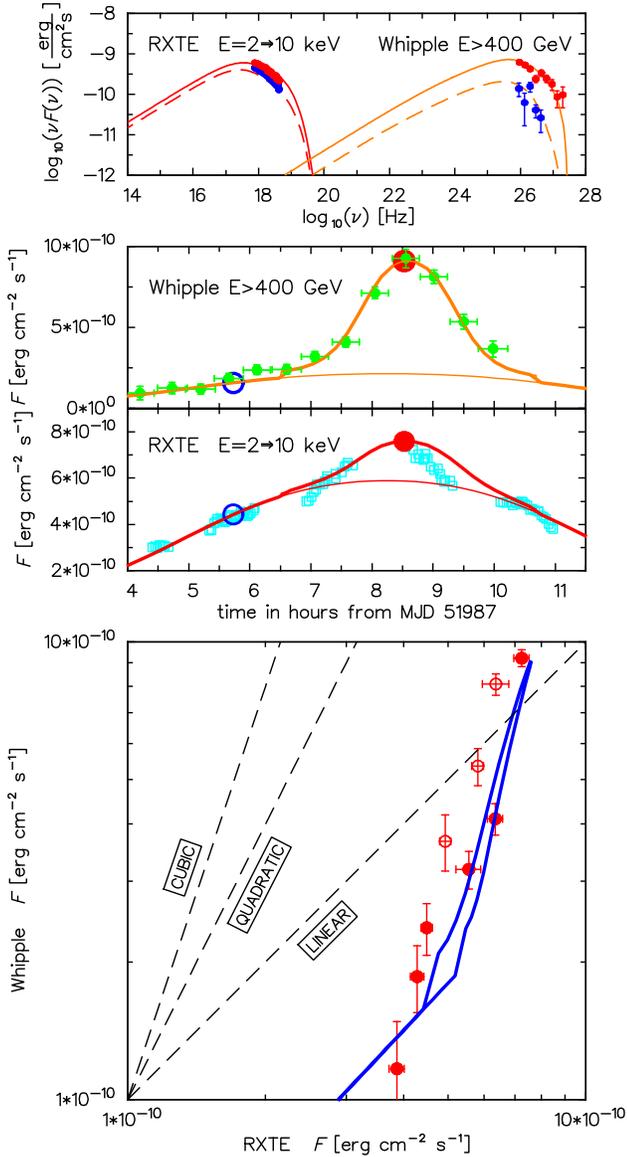}}
\caption{The same activity of Mrk 421 as in Fig. \ref{fig_inj} and the results
         of our modelling which assume the evolution of the Doppler factor. Note that
         in this particular case the correlation for a single source is perfectly
         linear. This is well visible in the bottom part of the correlation 
         panel, where only the background source is contributing to the total 
         emission.
        }
\label{fig_beam}        
\end{figure}

An intrinsically isotropic emission of a source that travels with the 
relativistic velocity $V = \beta c$ is confined in a beam along the velocity vector. 
The half-opening angle of the beam is $\phi \simeq 1/\Gamma$ in radians, where 
$\Gamma=1/\sqrt{1-\beta^2}$ is the Lorentz factor. This is the well-know 
beaming effect that amplifies the observed emission. For a given
angle $\theta$ between the velocity and the direction to the 
observer the amplification of the emission $F_{\nu} = \delta^3 F'_{\nu'}$
is described by the Doppler factor $\delta = 1/(\Gamma(1 - \beta \cos \theta))$.
Therefore a small increase or decrease of $\theta$ may cause 
significant variations of the observed flux. This may happen 
when the source travels on a helical or quasi-helical trajectory.
Such trajectories of the jets were observed many times in
AGNs and may cause quasi-periodic activity observed in blazars
(e.g. Rani et al. \cite{Rani09}, Qian et al. \cite{Qian09}).
Still for the sake of simplicity we do not describe
the trajectory of the source in our simulations. This will require 
many free parameters and is not necessary to test our assumptions.
We simply describe the change of the viewing angle by the cosine
function
\begin{equation}
\theta (y) = \theta_{\rm max} - \Delta \theta \frac{\cos(y) + 1}{2},
\end{equation}
where $\theta_{\rm max} = 180 /(\pi \Gamma)$ and $\Delta \theta$ is
the amplitude of the change. Parameter $y$ changes linearly from 
$-\pi$ to $\pi$ during a flare, where $t_a$ describes the duration 
of the activity in the observer's frame. 

Note that the amplification of the emission is
frequency-independent and will always give a linear 
correlation ($x_{\uparrow} = x_{\downarrow} = 1$). This is the
main advantage of this approach. Moreover, the correlation slope 
does not depend on the evolution of the Doppler factor in time. 
Therefore, the activity can be simulated for example by the 
source acceleration and  deceleration, but this approach 
requires special justification. We assume a constant value 
of the source velocity and small changes of the viewing angle 
that give significant variations of the observed flux. Note 
that this requires a relatively small value of $\theta < 10^o$ 
in general.

To describe the observed spectra we use a very simple model of a  
spherical and homogeneous source with a uniform density and 
magnetic field. The approach used before which assumes simultaneous 
emission of a big and small source is also used here to explain the 
more than linear correlation. We assume the same velocity for both 
sources that gives $\Gamma = 10$, and $\theta_{\rm max} = 
5.73$ deg and the same amplitude of the viewing angle 
change $\Delta \theta = 4.73$. However, the assumed duration 
of the activity for the big source is $t_a=18.5$ h, whereas for 
the small one it is $t_a = 4.2$ h in the observer's frame. 
This indicates that the sources are traveling along different
trajectories. The difference appears also in 
the radius $R = 10^{16}$ and $1.45 \times 10^{15}$ cm, 
the magnetic field strength $B = 0.1$ and $0.05$ G and 
the particle density $K = 2.5 \times 10^{3}$ and 
$2 \times 10^{5}$ cm$^{-3}$ for the big and small source
respectively. The particle energy distribution inside the 
big source is assumed to be a broken power-law function 
$N(\gamma) = K \gamma^{-n_1}$ for $\gamma < \gamma_{\rm brk}$
and $N(\gamma) = K \gamma_{\rm brk}^{n_2 - n_1} \gamma^{-n_2}$
above the break. 
We use $\gamma_{\rm min} = 1$, $\gamma_{\rm brk} = 2 
\times 10^5$, $\gamma_{\rm max} = 10^6$, $n_1=2$ and 
$n_2 = 4$ to describe this spectrum. The energy spectrum
inside the small source is approximated by a power-law function, 
where we use  $\gamma_{\rm min} = 1$, $\gamma_{\rm max} = 2 
\times 10^6$ and $n=2$. The values used here are very similar
to the parameters used in the injection scenario. Note that the
intrinsic emission of the sources is constant in time. This
means that the sources are in equilibrium, where the
radiative cooling is fully compensated by the acceleration.
The probability of such a situation is rather low, but
this scenario is physically possible. Dominance of the
acceleration or the cooling will slightly disturb the 
perfect linear correlation provided by the Doppler boosting effect.
However, to test an ideal case, where for a single source
$x_{\uparrow} = x_{\downarrow}$, we assume the equilibrium.

To compare the boosting scenario with the approach 
presented before we simulate the same activity of Mrk 421.
The result of the simulation is presented in Fig. \ref{fig_beam}. 
This particular model provides a better fit for the observed
spectra and the light curves. However, this is a very simple
scenario that has more free parameters. A single source emission
provides a perfectly linear correlation which is well visible in
the lower part of the correlation panel in Fig. \ref{fig_beam},
where only the big source emission is correlated. 
Simultaneous emission of both sources gives almost an cubic 
correlation. However, the correlation is not the same for 
the rise and decay of the flare. This is the result of the 
relative shift in time between the light curves produced 
by each source. The flare produced by a single source is
symmetric in time in this particular scenario. But
the maximum of the flare produced by the small source is 
delayed by about half an hour in comparison to the maximum
of the big source activity. This causes the small difference 
between $x_{\uparrow}$ and $x_{\downarrow}$. This shows an
important fact: that the correlation produced by many 
independent sources can be diluted simply by the delays 
between the flares. This may be the reason why 
the correlation was well determined only in a few cases. Finally, 
the calculated correlation is slightly shifted in comparison with the observed 
correlation. This is the result of the delay between the observed 
X-ray and gamma-ray flare. Such a delay cannot be simulated in 
our simple scenario.

\section{Summary}

Single zone models frequently used to explain emission of blazars 
in the TeV range are not able to explain quadratic or more than 
quadratic correlation between the X-ray and gamma-ray emission.
So we propose a simple solution to this problem which
assumes a simultaneous emission of at least two independent 
sources. In the first approach we use the model that simulates
activity by the particle injection. This rather classical
approach was successfully applied in many similar models. But
our calculations show that the simulated correlation has 
different slopes for the rising and decay phase of the flare.
This seems to be a general problem for the models that assume 
different processes (e.g. acceleration and cooling) to explain 
rise and decay of a flare.
This lead us to propose an alternative approach, where the activity
is produced by the Doppler boosting effect. This scenario 
provides a perfectly linear correlation for the single source
emission in the rising and decaying phase of the activity. 
But we still have to simulate simultaneous emission of 
the two sources to obtain quadratic or cubic correlation. 
The proposed approach has several advantages:

\begin{itemize}
\item{it can explain any slope of the correlation,}
\item{in the extreme case it is possible to explain the orphan flares,} 
\item{the approach does not involve a new model of the emission, 
      it uses the standard SSC scenario to explain a single source 
      radiation,}
\item{it may explain why the correlation was well determined only 
      in a few cases so far,}      
\item{in was already shown that using this approach it is possible to
      explain also the rapid variability.}      
\end{itemize}

The correlation may give important information about the region 
of the jet where the high energy activity is generated. However, 
many further observations are required to fully understand the nature 
of the correlation. There are still many open questions. For example, 
the correlation is observed exceptionally, is this normal for 
TeV blazars or is this rather a problem with the observation 
quality? Is there any preferable slope of the correlation?
How does the correlation look in different sources? Answers for
the above questions and for many others will require precise,
simultaneous observations in the X-ray and gamma-ray
range.


\begin{thebibliography}{}
\bibitem[2009]{Aharonian09}   Aharonian, F., Akhperjanian, A. G., Anton, G., et al.        2009, arXiv:0906.2002      
\bibitem[2007]{Aharonian07}   Aharonian, F., Akhperjanian, A. G., Bazer-Bachi, A. R., et al. 2007, ApJ, 664, 71
\bibitem[2007a]{Albert07a}    Albert, J., Aliu, E., Anderhub, H., et al.                   2007a, ApJ, 669, 1143
\bibitem[2007b]{Albert07b}    Albert, J., Aliu, E., Anderhub, H., et al.                   2007b, ApJ, 663, 125
\bibitem[2005]{Blazejowski05} Blazejowski, M., Blaylock, G., Bond, I. H.,  et al.          2005, ApJ, 630, 130
\bibitem[1996]{Bloom96}       Bloom, S. D. \& Marscher, A. P.                              1996, ApJ, 461, 657
\bibitem[2009]{Bonnoli09}     Bonnoli, G., Ching-Cheng Hsu, Goebel, F., et al.             2009, arXiv:0907.0831v1
\bibitem[1997]{Catanese97}    Catanese, M., Bradbury, S., M., Breslin, A., C., et al.      1997, ApJ   , 487, L143
\bibitem[1999]{Chiaberge99}   Chiaberge, M. \& Ghisellini, G.,                             1999, MNRAS, 306, 551
\bibitem[1999]{Coppi99}       Coppi, P. S. \& Aharonian, F. A.,                            1999, ApJ, 521, 33
\bibitem[1997]{Dermer97}      Dermer, C. D., Sturner, S. J. \& Schlickeiser, R.,           1997, ApJS, 109, 103
\bibitem[2009]{Donnarumma09}  Donnarumma, I., Vittorini, V., Vercellone, S., et al.        2009, ApJ, 691, 13
\bibitem[2008]{Fossati08}     Fossati, G., Buckley, J. H., Bond, I. H., et al.             2008, ApJ, 677, 906
\bibitem[2006]{Gliozzi06}     Gliozzi, M., Sambruna, R. M., Jung, I., et al.               2006, ApJ, 646, 61
\bibitem[1996]{Ghisellini96}  Ghisellini, G., Maraschi, L., Dondi, L.,                     1996, A\&ASS, 120, 503
\bibitem[1979]{Gould79}       Gould R.J,                                                   1979, A\&A, 76, 306
\bibitem[2008]{Graff08}       Graff, P., B., Georganopoulos, M., Perlman, E., S., et al.   2008, ApJ, 689, 68G
\bibitem[2009]{Horan09}       Horan, D., Acciari, V.A., Bradbury, S.M., et al.             2009, arXiv:0901.1225v1      
\bibitem[1996]{Inoue96}       Inoue, S. \& Takahara, F.,                                   1996, ApJ, 463, 555
\bibitem[1997]{Mastichiadis97}Mastichiadis, A. \& Kirk, J. G.,                             1997, A\&A, 320, 19
\bibitem[2000]{Kataoka00}     Kataoka, J., Takahashi, T., Makino, F., et al.               2000, ApJ, 528, 243
\bibitem[2001]{Katarzynski01} Katarzynski, K., Sol, H., Kus, A.,                           2001, A\&A, 367, 809
\bibitem[2005]{Katarzynski05} Katarzynski, K., Ghisellini, G., Tavecchio, F., et al.       2005, A\&A, 433, 479
\bibitem[2006]{Katarzynski06} Katarzynski, K., Ghisellini, G., Mastichiadis, A., et al.    2006, A\&A, 453, 47 
\bibitem[2008]{Katarzynski08} Katarzynski, K., Lenain, J.-P., Zech, A., et al.             2008, 2008, MNRAS, 390, 371
\bibitem[1998]{Kirk98}        Kirk, J. G., Rieger, F. M. \& Mastichiadis, A.,              1998, A$\&$A, 333, 452
\bibitem[2000]{Krawczynski00} Krawczynski, H., Coppi, P. S., Maccarone, T., Aharonian, F. A., 2000, A\&A, 353, 97
\bibitem[2002]{Krawczynski02} Krawczynski, H., Coppi, P. S., \& Aharonian, F.,             2002, MNRAS, 336, 721
\bibitem[2004]{Krawczynski04} Krawczynski, H., Hughes, S. B., Horan, D., et al.            2004, ApJ, 601, 151
\bibitem[1998]{Pian98}        Pian, E., Vacanti, G., Tagliaferri, G., et al.,              1998, ApJ, 492, L17
\bibitem[2009]{Qian09}        Qian, Shan-Jie, Witzel, A., Zensus, J. A., et al.            2009, RAA, 9, 137
\bibitem[2004]{Sokolov04}     Sokolov, A., Marscher, A. P. \& McHardy, I. M.,              2004, ApJ, 613, 725
\bibitem[2000]{Sambruna00}    Sambruna, R. M., Aharonian, F. A., Krawczynski, H., et al.   2000, ApJ, 538, 127
\bibitem[2000]{Takahashi00}   Takahashi, T.; Kataoka, J.; Madejski, G., et al.             2000, ApJ, 542, 105
\bibitem[2009]{Rani09}        Rani, B., Wiita, P., J., \& Gupta, A., C.,                   2009, ApJ, 696, 2170
\bibitem[1978]{Rees78}        Rees, M.J.,                                                  1978, MNRAS, 184, 61
\bibitem[1999]{Villata99}     Villata, M., \& Raiteri, C. M.,                              1999, A\&A, 347, 30
\end{thebibliography}
\end{document}